\documentclass[twocolumn,showpacs,preprintnumbers,amsmath,amssymb,pre]{revtex4-1}

\usepackage{graphicx}

\begin{document}

\title{Self-avoiding walks and connective constants in clustered 
       scale-free networks}
\author{Carlos P. Herrero}
\affiliation{Instituto de Ciencia de Materiales,
         Consejo Superior de Investigaciones Cient\'ificas (CSIC),
         Campus de Cantoblanco, 28049 Madrid, Spain}
\date{\today}

\begin{abstract}
Various types of walks on complex networks have been used in recent years to
model search and navigation in several kinds of systems, with particular
emphasis on random walks. This gives valuable information on network properties,
but self-avoiding walks (SAWs) may be more suitable than unrestricted
random walks to study long-distance characteristics of complex systems. 
Here we study SAWs in clustered scale-free networks, characterized by a 
degree distribution of the form $P(k) \sim k^{-\gamma}$ for large $k$.
Clustering is introduced in these networks by inserting three-node loops
(triangles).  The long-distance behavior of SAWs gives us information on 
asymptotic characteristics of such networks.
The number of self-avoiding walks, $a_n$, has been obtained by direct 
enumeration, allowing us to determine the {\em connective constant}
$\mu$ of these networks as the large-$n$ limit of the ratio $a_n / a_{n-1}$.
An analytical approach is presented to account for the results derived from
walk enumeration, and both methods give results agreeing with each other.
In general, the average number of SAWs $a_n$ is larger for clustered networks
than for unclustered ones with the same degree distribution. 
The asymptotic limit of the connective constant for large system size $N$ 
depends on the exponent $\gamma$ of the degree distribution:
For $\gamma > 3$, $\mu$ converges to a finite value as $N \to \infty$;
for $\gamma = 3$, the size-dependent $\mu_N$ diverges as $\ln N$,
and for $\gamma < 3$ we have $\mu_N \sim  N^{(3 - \gamma) / 2}$.
\end{abstract}

\pacs{89.75.Hc, 05.40.Fb, 87.23.Ge, 89.75.Da}
%

\maketitle

\section{Introduction}

In last decades, research in various fields has shown evidence that
many types of real-life systems can be described in terms of
networks, where nodes represent typical system units and edges
correspond to interactions between connected pairs of units. Such topological
characterization has been applied to describe natural and artificial
systems, and is used at present to analyze processes
occurring in real systems (social, economic, technological, biological)
\cite{do03,co10,al02,st01,fe13}.

Several kinds of theoretical and experimental techniques have been
employed to study and characterize a diversity of networks \cite{co07,ne10}. 
Various of these methods aim at studying dynamical processes, 
such as spread of infections \cite{mo00,ku01,he10,ba12,li17},
signal propagation \cite{wa98,he02b}, and
random spreading of information and opinion \cite{pa01,mo04,ca07b}.
The network structure plays an important role in these processes, 
as has been shown by using stochastic dynamics and 
random walks \cite{ja01,al03,ne03,no04,co07b,lo12}.

It turns out that many systems can be described by so-called scale-free (SF) 
networks, displaying a power-law distribution of degrees.
In an SF network the degree distribution $P_{\rm sf}(k)$, where $k$ is the 
number of links connected to a node, has a power-law decay
$P_{\rm sf}(k) \sim k^{-\gamma}$.
Networks displaying such a degree distribution have been found in both
natural and artificial systems, e.g., internet \cite{si03},
world-wide web \cite{al99}, social systems \cite{ne01}, 
and protein interactions \cite{je01}.
In these networks, the exponent $\gamma$ describing the distribution 
$P_{\rm sf}(k)$ has been usually found in the range 
$2 < \gamma < 3$ \cite{do03,go02}.
SF networks have been used to study statistical 
physics problems, as avalanche dynamics \cite{go03b}, 
percolation \cite{ra09}, and 
cooperative phenomena \cite{do02b,he04,ba06,do08,he09,do10,os11}. 

Self-avoiding walks (SAWs) can be more effective than unrestricted random 
walks in exploring networks, since they are not allowed to return to sites 
already visited in the same walk \cite{ki16,ti16}. This property has been used
to define local search strategies in scale-free networks \cite{ad01}.
SAWs have been employed with various purposes, such as modeling structural 
and dynamical aspects of polymers \cite{ge79,or07,be12,gu14}, conformation of DNA 
molecules \cite{ma99,wi08},
characterization of complex crystal structures \cite{he95,he14b},
and analysis of critical phenomena in lattice models \cite{kr82,cl10,gu13}. 
In the context of complex networks, several features of SAWs 
have been studied in small-world \cite{he03},
scale-free \cite{he05a}, and fractal networks \cite{ho14,fr14}.

The asymptotic properties of SAWs in regular and complex networks are
usually studied in connection with the so-called {\em connective constant} or
long-distance effective connectivity, which quantifies the increase in the
number of SAWs at long distances \cite{ra85,pr91}. 
For SF networks, in particular, this has allowed to distinguish 
different regimes depending on the exponent $\gamma$
of the distribution $P_{\rm sf}(k)$ \cite{he05a}.
One can also consider kinetic-growth self-avoiding walks on complex networks,
to study the influence of {\em attrition} on the maximum length of 
the paths \cite{he05b,he07}, but this kind of walks will not be addressed here.

Many real-life networks include clustering, i.e., the probability of
finding loops of small size is larger than in random networks.
This has been in particular quantified by the so-called clustering
coefficient, which measures the likelihood of three-node loops
({\em triangles}) in a network \cite{ne10}.
The relevance of loops for different aspects of networks is now generally
recognized, and several models of clustered networks have been defined and 
analyzed by several research groups \cite{ho02,kl02,se05,kl18,lo18}. 
In recent years, it has been shown that generalized random graphs can be
generated incorporating clustering in such a way that exact formulas can
be derived for many of their properties \cite{ne09,mi09,he11}.
This includes the study of physical problems such as critical phenomena in
SF networks, e.g. the Ising model \cite{he15}.
For an exponent $\gamma \leq 3$ it was found that clustered and unclustered
networks with the same size and degree distribution $P(k)$ have different
paramagnetic-ferromagnetic transition temperature $T_c$, what indicates that
clustering favors ferromagnetic correlations and causes an increase in $T_c$.
Other works on clustered networks have addressed different questions, 
as robustness \cite{hu13}, bond percolation \cite{gl09,gl10,ra16},
and spread of diseases \cite{mo12,wa12}.

Here we study long-range properties of SAWs in clustered SF networks.
We pose the question whether clustering significantly changes the properties
of SAWs in this kind of networks. This is particularly interesting for
the long-distance behavior of SAWs, which is expected to depend on
network characteristics such as cluster concentration and decay of the
degree distribution for large degree (exponent $\gamma$ for SF networks).
Thus, we focus on the influence of introducing clusters (here triangles) 
upon the asymptotic limits of SAWs and connective constants.
The average number of walks for a given length $n$ is calculated
by an iterative procedure, and the results are compared with those obtained 
from direct enumeration in simulated networks.
Both methods yield results which agree with each other in the different
regions defined by the exponent $\gamma$ and for a wide range of cluster 
densities. Comparing results for clustered and unclustered networks with 
the same degree distribution, we find for large networks with $\gamma > 3$
that the long-distance behavior of SAWs is not affected by clustering.
On the other hand, clustering changes the connective constants
derived from SAWs for networks with $\gamma \le 3$.

The paper is organized as follows. 
In Sec.\,II we describe the clustered SF networks studied here. 
In Sec.\,III we present some generalities on SAWs and its application to
unclustered scale-free networks.  In Sec.\,IV we introduce an analytical
method employed to calculate the number of SAWs in clustered networks,
and in Sec.\,V we compare results of this analytical procedure with those
obtained by directly enumerating SAWs in simulated networks.
The paper closes with the conclusions in Sec.\,VI.

\section{Description of the networks}

\subsection{Networks construction}

We study here clustered networks with a degree distribution $P(k)$,
which for large degree $k$ follows a power-law 
$P_{\rm sf}(k) \sim k^{-\gamma}$. The exponent controlling the decay
of the distribution is taken as $\gamma > 2$, so that the mean degree 
$\langle k \rangle$ remains finite in the large-size limit.
Clustering is introduced by including triangles into the
networks, i.e., triads of connected nodes.
One can consider other types of polygons (squares, pentagons, ...)
to study their effect on the properties of clustered networks, but
we take triangles as they are expected to give rise to stronger 
correlations between entities defined on network sites \cite{he15}.
The analytical method described here to study SAWs in the presence of
triangles can be easily extended to other types of motifs.

\begin{figure}
\vspace{-0.0cm}
\includegraphics[width=7.0cm]{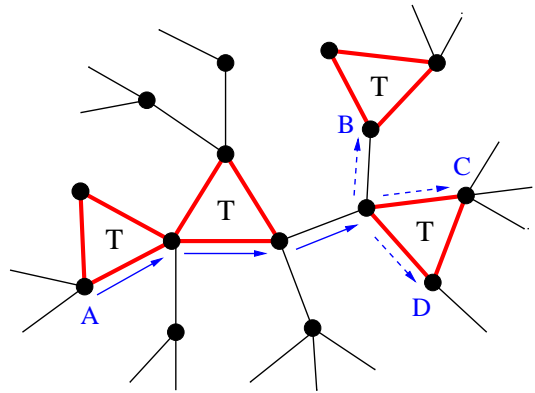}
\vspace{-0.2cm}
\caption{
Schematic representation of a typical network considered in this
work, for which one separately specifies the number of single edges
and triangles (bold red edges) attached to each node.
Triangles are indicated as T. Arrows indicate a possible SAW
starting from node A.
For the fourth step ($n = 4$), there are three available edges
(one $s$-link and two $t$-links) reaching nodes B, C, and D.
}
\label{f1}
\end{figure}

Our networks are generated by using the procedure described by 
Newman \cite{ne09}, where the number of single edges and the number of 
triangles are independently defined.
A method like this permits us to manage generalized random graphs, which
incorporate clustering in a rather simple manner, 
thus allowing one to analytically study various properties of the resulting 
networks \cite{ne09}.
Given a network, we call $N$ the number of nodes, $t_i$ the number of
triangles of which node $i$ is a vertex, and $s_i$ the number of single links
not included in triangles ($i = 1, ..., N$). 
This means that, for the purpose of network construction, edges within 
triangles are considered apart from single links.
Then, a single link can be regarded as an element connecting two nodes and 
a triangle as a network component joining together three nodes.
Thus, the degree $k_i$ of node $i$ is given by $k_i = s_i + 2 \, t_i$, since
each triangle connects it to two other nodes.
A schematic plot of this kind of networks is displayed in Fig.~1, where
triangles are marked as T. In this figure, single links appear as
black lines and edges belonging to triangles are depicted as bold red lines.
For clarity of the presentation below, both kinds of edges will be denoted 
$s$-links and $t$-links, respectively.

The networks are built in two steps. In the first one, we introduce 
the edges by connecting pairs of nodes.
We ascribe to each node $i$ a random integer $s_i$, which will be 
the number of outgoing links (stubs) from this node. 
The numbers $\{ s_i \}_{i=1}^N$ are picked up from the probability 
distribution $P_{\rm sf}(s) \sim s^{-\gamma}$, assuming that 
$s_i \geq k_0$, the minimum allowed degree \cite{ne05}. 
This gives a total number of stubs $K = \sum_{i=1}^N s_i$, which 
we impose to be an even integer.
Once the numbers $s_i$ are defined, we connect stubs at random
(giving a total of $L = K/2$ connections), with the conditions:
(i) no two nodes can have more than one edge connecting them
   (no multiedges), and
(ii) no node can be connected by a link to itself (no self-edges).
Networks fulfilling these conditions are usually called simple networks
or simple graphs \cite{ne10}. 

In a second step we incorporate $N_{\Delta}$ triangles into the considered 
network. $N_{\Delta}$ is defined by the parameter $\nu$, 
the mean number of triangles in which a node is included,
$N_{\Delta} = \frac13 N \nu$.
The number of triangles associated to each node is taken from
a Poisson distribution $Q(t) = {\rm e}^{-\nu} \nu^t / t!$.
This gives us $t_i$ {\em corners} corresponding to node $i$, and the total
number is $T = \sum_{i=1}^N t_i = 3 N_{\Delta}$.
For consistency, $T$ has to be a multiple of 3.
Then, we randomly choose triads of corners to form triangles, 
avoiding multiple and self-edges as in conditions (i) and (ii) 
in the previous paragraph.

\begin{figure}
\vspace{-0.9cm}
\includegraphics[width=8.0cm]{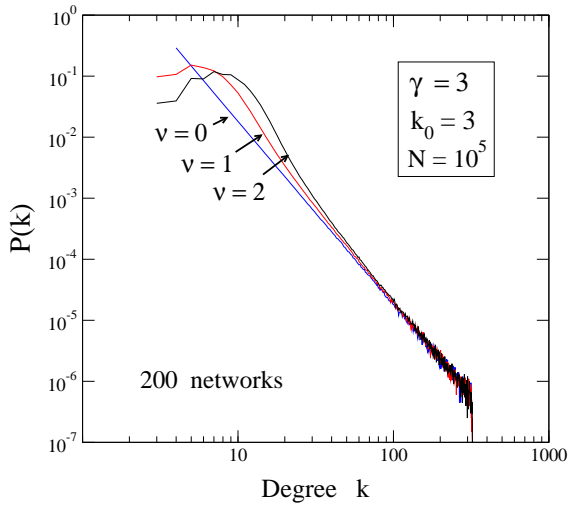}
\vspace{-0.5cm}
\caption{
Probability density $P(k)$ as a function of the degree $k$ for networks
with $\gamma = 3$, minimum degree $k_0$ = 3, and $N = 10^5$ nodes.
The displayed data are an average over 200 network realizations for each
value of the parameter $\nu$ = 0, 1, and 2.
}
\label{f2}
\end{figure}

It is known that long-tailed distributions including nodes with very
large degree may display undesired correlations between the degrees 
of adjacent nodes, especially for exponents $\gamma < 3$ \cite{ca05b}.
For this reason we have introduced in the networks considered here 
a maximum-degree cutoff $s_{\rm cut} = \sqrt{N}$, which avoids such
correlations (see Sec.~II.B) \cite{ca05b,he15}.
This restriction is in fact only effective for $\gamma < 3$, as
explained below.

The degree distribution $P(k)$ obtained for clustered networks 
generated by following the procedure presented above is shown in Fig.~2.
In this figure, we display $P(k)$ for networks with $N = 10^5$ nodes, 
$\gamma = 3$ and $k_0 = 3$. Shown are results for three values of the
triangle density: $\nu$ = 0, 1, and 2, corresponding in each
case to an average over 200 network realizations. One observes that
increasing the value of $\nu$ causes clear changes in $P(k)$ for 
small $k$ values, with respect to the degree distribution for $\nu = 0$.
For large degrees, however, the distribution follows a dependence 
$P_{\rm sf}(k) \sim k^{-\gamma}$ characteristic of 
scale-free networks with $\gamma = 3$.
This could be expected from the fact that the Poisson distribution 
$Q(t)$ associated to the triangles has a fast exponential-like decay
for large $t$.

In summary, our networks are defined by the following parameters:
$N$ (number of nodes), $k_0$ (minimum degree), $\gamma$ (exponent
controlling the distribution of single edges, $P_{\rm sf}(s)$),
and $\nu$ (density of triangles). 
For the numerical simulations we have generated networks with several
values of these parameters.
For each set ($N$, $k_0$, $\gamma$, $\nu$), we considered
different network realizations, and for a given network we selected
at random the starting nodes for the SAWs.
For each considered parameter set, the total number of generated SAWs
amounted to about $2 \times 10^6$.
All networks considered here contain a single component, i.e. any node in
a network can be reached from any other node by traveling through a finite
number of links.

For comparison with the results obtained for clustered networks, we
have also generated networks with the same degree distribution $P(k)$
than the clustered ones, but without explicitly including triangles.
This means that these networks are built up from the degree sequence
$\{ k_i \}_{i=1,...,N}$ given by $k_i = s_i + 2 t_i$, but randomly 
connecting the $k_i$ outgoing links (stubs) for each node $i$ as
indicated above for $s$-links. This corresponds to the so-called
{\em configuration model} \cite{ne10}.

\subsection{Mean values $\langle k \rangle$ and $\langle k^2 \rangle$}

Important characteristics of the considered networks, which will be
used below in our calculations, are the
mean values $\langle k \rangle$ and $\langle k^2 \rangle$.
For scale-free networks with $\nu = 0$ (no clustering), the average 
degree is given by
\begin{equation}
 \langle s \rangle_{\infty} = \sum_{s=k_0}^{\infty} s \, P_{\rm sf}(s)
        \approx   k_0 \,  \frac{\gamma-1}{\gamma-2}   \; ,
\label{kmeaninf}
\end{equation}
where the expression on the right has been obtained by replacing the sum by 
an integral, which is justified for large $N$.
For our networks including triangles ($\nu > 0$), we have 
$k_i = s_i + 2 t_i$, and
\begin{equation}
  \langle k \rangle_{\infty}  =  \langle s \rangle_{\infty}  +  2 \, \nu 
       \approx   k_0 \, \frac{\gamma-1}{\gamma-2}  +  2 \, \nu    \; .
\label{kmean}
\end{equation}

For clarity of the presentation we write $\langle s \rangle$ to indicate 
an average value for unclustered SF networks (configuration model), i.e. 
consisting of $s$-links. We write $\langle k \rangle$ in the general case,
which includes clustered networks. Moreover, the subscripts $N$ and $\infty$
refer to networks of size $N$ and to the infinite-size limit (when it exists),
respectively. When no subscript appears, we understand that it refers to a
general case, without mention to the system size.

For finite networks, a size effect appears in the mean degree, 
as a consequence of the effective cutoff $k_{\rm cut}$ appearing in 
the degree distribution. In fact, for a given network of size $N$ 
one has \cite{do02b,ig02}
\begin{equation}
   \sum_{k_{\rm cut}}^{\infty} P_{\rm sf}(s) = \frac{c}{N}  \; ,
\label{sumkc}
\end{equation}
where $c$ is a constant on the order of unity. This yields
for $k_{\rm cut} \gg k_0$  \cite{he15}:
\begin{equation}
   k_{\rm cut} \approx k_0 \left( \frac{N}{c} \right)^{\frac{1}{\gamma-1}} \, ,
\label{kcutk}
\end{equation}
so that $k_{\rm cut} \sim \, N^{1/(\gamma-1)}$, as in \cite{do02b,ig02}.
From Eq.~(\ref{kcutk}), one obtains for finite scale-free networks \cite{he15}:
\begin{equation}
     \langle s \rangle_N \approx \langle s \rangle_{\infty}
         \left[ 1 - \left( \frac{c}{N} \right)^{\frac{\gamma-2}{\gamma-1}}
            + {\cal O} \left( \frac1N \right)   \right]   \; .
\label{smean_app}
\end{equation}

For the mean value $\langle s^2 \rangle$, the size dependence and
its large-size behavior change with the exponent $\gamma$.
For SF networks with $\gamma > 3$, the dependence of $\langle s^2 \rangle$
on $N$ is similar to that of $\langle s \rangle$, namely \cite{he15} :
\begin{equation}
   \langle s^2 \rangle_N \approx \langle s^2 \rangle_{\infty}
        \left[ 1 - \left( \frac{c}{N} \right)^{\frac{\gamma-3}{\gamma-1}}
          + {\cal O} \left( \frac1N \right)   \right]  \; ,
\label{smean2_app}
\end{equation}
with $\langle s^2 \rangle_{\infty} \approx k_0^2 \, (\gamma-1) / (\gamma-3)$.

For $\gamma = 3$, we have
\begin{equation}
    \langle s^2 \rangle_N  =   \sum_{k_0}^{k_{\rm cut}} s^2 P_{\rm sf}(s)
       \approx  \frac{2}{k_0^{-2} - k_{\rm cut}^{-2}}  \,
               \ln \frac{k_{\rm cut}}{k_0}  \; ,
\label{smean2_app2}
\end{equation}
and using Eq.~(\ref{kcutk}) we find for $k_{\rm cut} \gg k_0$:
\begin{equation}
  \langle s^2 \rangle_N =  k_0^2 \, \ln N + {\cal O} ( 1 )    \; .
\label{smean2_app3}
\end{equation}

For SF networks with an exponent $\gamma < 3$,
Catanzaro {\em et al.} \cite{ca05b} found appreciable correlations
between degrees of adjacent nodes when no multiple and
self-edges are allowed. These degree correlations can be avoided
by introducing a cutoff $k_{\rm cut} \sim N^{1/2}$, as indicated above.
Thus, for $\gamma < 3$ we generate here networks with a cutoff
$k_{\rm cut} = N^{1/2}$.
Note that for $\gamma > 3$ the cutoff derived above from the condition 
given in Eq.~(\ref{sumkc}) is more restrictive than putting 
$k_{\rm cut} = N^{1/2}$, so that in this case the effective cutoff
is given by Eq.~(\ref{kcutk}).
Then, for finite networks with $\gamma < 3$ the mean values
$\langle s \rangle_N$ and $\langle s^2 \rangle_N$ are given by:
\begin{equation}
  \langle s \rangle_N  \approx  \langle s \rangle_{\infty}
       \left[ 1 - \left( \frac{k_0}{\sqrt{N}} \right)^{\gamma-2}  \right]  \; ,
\label{smean_app2}
\end{equation}
and
\begin{equation}
   \langle s^2 \rangle_N \approx  k_0^2  \, \frac{\gamma - 1}{3 - \gamma}
         \left( \frac{\sqrt{N}}{k_0} \right)^{3 - \gamma}   \; .
\label{smean2_app4}
\end{equation}

For clustered SF networks with a triangle density $\nu > 0$,
the average value $\langle k^2 \rangle$  is given by
\begin{equation}
  \langle k^2 \rangle =  \langle (s + 2 t)^2  \rangle =
           \langle s^2 \rangle + 4 \langle s \rangle \langle t \rangle +
           4 \langle t^2 \rangle   \; ,
\label{k2s2}
\end{equation}
where we have used the fact that 
$\langle s \, t \rangle = \langle s \rangle  \langle t \rangle$,
since variables $s$ and $t$ are independent for the way 
of building up these networks.
Then, for clustered networks with $\gamma > 3$, we have in the
large-size limit:
\begin{equation}
 \langle k^2 \rangle_{\infty}  \approx  k_0^2 \, \frac{\gamma - 1}{\gamma - 3} +
        4 \, k_0 \, \nu \, \frac{\gamma - 1}{\gamma - 2} +
        4 \, \nu \, (\nu + 1)      \, ,
\label{k2mean}
\end{equation}
where $\nu$ and $\nu \, (\nu + 1)$ are the average values
$\langle t \rangle$ and $\langle t^2 \rangle$ corresponding to the
Poisson distribution of triangles $Q(t)$.
For scale-free networks with $\gamma \leq 3$ one can write expressions
for $\langle k^2 \rangle_N$
derived from Eq.~(\ref{k2s2}) by using the corresponding formulas for 
$\langle s \rangle_N$ and $\langle s^2 \rangle_N$ given above.

\subsection{Clustering coefficient}

The clustering coefficient $C$ is usually defined for complex networks 
as the ratio \cite{ne10,ne09,yo11}
\begin{equation}
   C = \frac{3 N_{\Delta}}{N_3}   \; ,
\end{equation}
where $N_{\Delta}$ is the number of triangles and $N_3$ is the
number of connected triplets. Here a connected triplet means three
nodes $n_1 n_2 n_3$ with links ($n_1, n_2$) and ($n_2, n_3$), and
$N_3$ is given by
\begin{equation}
  N_3 = N \sum_k \frac{k (k-1)}{2} P(k) =
      \frac12 N \left( \langle k^2 \rangle - \langle k \rangle  \right) \; .
\end{equation}
Taking into account that the triangle density in our clustered networks
is related to the number of triangles $N_{\Delta}$ by the expression
$N_{\Delta} = N \nu / 3$, we have
\begin{equation}
    C = \frac {2 \nu}{ \langle k^2 \rangle - \langle k \rangle } \; .
\label{c_coeff}
\end{equation}

\begin{figure}
\vspace{-0.7cm}
\includegraphics[width= 8.0cm]{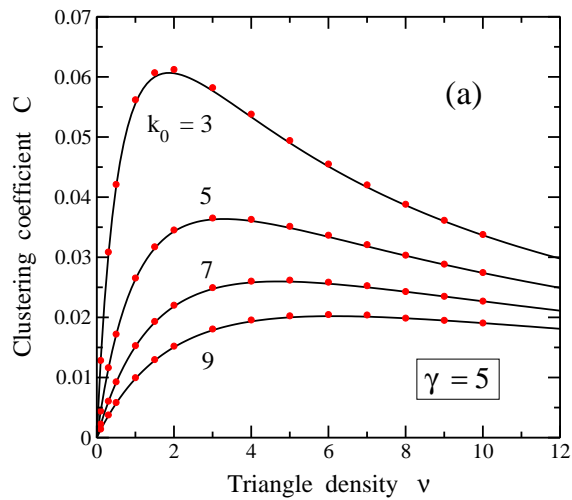}
~\vspace{-1.0cm}
\includegraphics[width= 8.0cm]{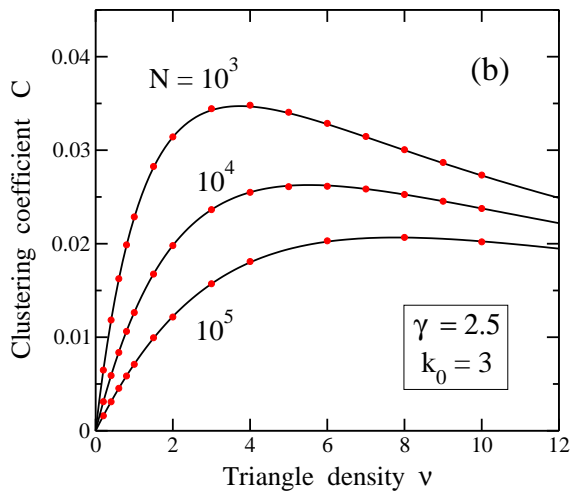}
\vspace{1.5cm}
\caption{
Clustering coefficient $C$ as a function of the triangle density $\nu$
for clustered scale-free networks.
(a) $\gamma = 5$ and various values of the minimum degree $k_0$.
From top to bottom, $k_0$ = 3, 5, 7, and 9.
Solid circles represent results derived from clustered networks with
$N = 10^5$ nodes. Solid lines were analytically derived in the limit
$N \to \infty$ by using the expressions given in the text.
(b) $\gamma = 2.5$ and three different network sizes: $N = 10^3$,
$10^4$, and $10^5$. Solid lines were calculated by using the formulas
for $\langle s \rangle_N$ and $\langle s^2 \rangle_N$ given in
Sec.~II.B for $\gamma < 3$.
}
\label{f3}
\end{figure}

Note that in our clustered networks it is possible for single edges to 
form triangles. 
The average number $N_{\Delta}^{\rm rd}$ of such triangles is 
given by \cite{ne10}
\begin{equation}
  N_{\Delta}^{\rm rd} = \frac16
     \left( \frac{\langle k^2 \rangle}{\langle k \rangle} - 1  \right)^3 \; .
\end{equation}
This mean value is small for scale-free networks with large $\gamma$,
and rises as $\gamma$ decreases.
The density of these triangles, $N_{\Delta}^{\rm rd}/N$, vanishes in the
thermodynamic limit for $\gamma > 7/3$, and for $\gamma < 7/3$ this ratio
scales as a power of $N$ with exponent $(7 - 3 \gamma) / 2$.
Calling $\nu_0$ the mean number of randomly-generated triangles per node, 
we have
\begin{equation}
   \nu_0 = \frac{3 N_{\Delta}^{\rm rd}}{N} \; ,
\end{equation}
which rapidly converges to zero for large $N$ for the network parameters
considered here. For the minimum value of $\gamma$ studied here,
$\gamma = 2.5$, we have $\nu_0 \approx \frac12 k_0^{3/2} N^{-1/4}$,
i.e., $\nu_0 \approx 0.15$ for $k_0 = 3$ and $N = 10^5$. 
With these parameters and $\gamma = 3$, we find $\nu_0 \approx 0.02$.

In Fig.~3 we show the clustering coefficient $C$ as a function of the
triangle density $\nu$ for clustered scale-free networks.
Panel (a) corresponds to $\gamma = 5$ and several values of the 
minimum degree $k_0$.
Solid symbols were derived from simulations of networks with $N  = 10^5$,
and the lines were obtained by using Eq.~(\ref{c_coeff}) with the
mean values $\langle k \rangle_{\infty}$ and $\langle k^2 \rangle_{\infty}$ 
given in Sec.~II.B, Eqs.~(\ref{kmean}) and (\ref{k2mean}).
Small differences between both sets of results are due to
the finite size of the simulated networks.
For small triangle density $\nu$, the clustering coefficient increases
for rising $\nu$ and reaches a maximum, which is more pronounced for
smaller $k_0$.
For a given triangle density $\nu$, $C$ decreases for increasing
minimum degree $k_0$, as a consequence of the rise in the difference 
$\langle k^2 \rangle - \langle k \rangle$  which appears 
in the denominator of Eq.~(\ref{c_coeff}).
 
In Fig.~3(b), $C$ is plotted vs $\nu$ for scale-free networks
with $\gamma = 2.5$ and three system sizes: $N = 10^3$, $10^4$, and
$10^5$.  Here one observes a clear dependence of $C$ on $N$, with the
clustering coefficient decreasing when rising $N$ for a given triangle
density $\nu$. For large $\nu$, one finds a slow decrease in $C$ for 
increasing $\nu$, with the different curves approaching one to the other.

\section{Generalities on self-avoiding walks}

A self-avoiding walk is defined as a walk along the edges of a network 
which cannot intersect itself. The walk is restricted to moving to 
a nearest-neighbor node in each step,
and the self-avoiding condition restricts the walk to visit only
nodes which have not been occupied earlier in the same walk.
Here we do not consider SAWs as kinetically-grown walks in a dynamical
process, and just calculate (i.e., count) the number of possible
SAWs starting from a given node in a given network.
This means that all those SAWs have the same weight for calculating
ensemble averages, e.g. the connective constant discussed below.
This is not the case of kinetically-grown SAWs, for which the weight
is in general not uniform \cite{he05a,he05b}.
The number of SAWs of length $n$ in complex networks
depends in general on the considered starting node, however some 
properties such as the connective constant of a given network
are independent of the initial node (see below).
In the following we will call $a_n$ the average number of SAWs of length $n$,
i.e. the mean value obtained by averaging over the
network sites and over different network realizations (for a given set 
of parameters $N$, $k_0$, $\gamma$, and $\nu$).

Self-avoiding walks have been traditionally studied on regular 
lattices. In this case, it is known that the number of SAWs increases
for large $n$ as $a_n \sim n^{\Gamma - 1} \mu^n$, 
where $\Gamma$ is a critical exponent
which depends on the lattice dimension $D$ and $\mu$ is 
the {\em connective constant} or effective coordination
number of the considered lattice \cite{ra85}.
For $D > 4$ one has $\Gamma = 1$ \cite{pr91,so95}.
For a lattice with connectivity $k_0$, the connective constant
verifies $\mu \le k_0 - 1$, and can be obtained from the large-$n$ limit
\begin{equation}
       \mu = \lim_{n\to\infty} \frac{a_n}{a_{n-1}}  \hspace{3mm} .
\label{mulim}
\end{equation}
This parameter depends on the particular topology of each
lattice, and has been calculated very accurately for two- and 
three-dimensional lattices \cite{so95}.

For complex networks in general the situation is rather different
than in the case of regular lattices in low dimensions. 
This is particularly clear for random networks, which
are locally tree-like and do not display the so-called attrition of
SAWs caused by the presence of small-size loops of connected nodes.
A simple case of network without loops is a Bethe lattice 
(or Cayley tree) with fixed connectivity $k_0$, for which
the number of SAWs is given by  $a_n^{BL} = k_0 (k_0-1)^{n-1}$,
and then $\mu_{BL} = k_0 - 1$.
For Erd\"os-R\'enyi random networks \cite{bo98} with poissonian
distribution of degrees, one has
$a_n^{\rm ER} = \langle k \rangle^n$ \cite{he03}, and then the connective
constant is $\mu^{\rm ER} = \langle k \rangle$.

For generalized random networks (configuration model), one has \cite{he05a}
\begin{equation}
a_n = \langle k \rangle  \left( \frac{\langle k^2 \rangle}{\langle k \rangle}
          - 1  \right)^{n-1}   \; .
\label{an1}
\end{equation}
In this expression, the ratio $\langle k^2 \rangle / \langle k \rangle$
is the average degree of a randomly-chosen end node of a randomly selected
edge \cite{do02,ne10}.
The term $-1$ in Eq.~(\ref{an1}) introduces the self-avoiding condition.
Thus, the connective constant $\mu$ for random networks is given by
\begin{equation}
   \mu = \frac{\langle k^2 \rangle}{\langle k \rangle} - 1  \; .
\label{mu_inf}
\end{equation}
Note that the expression given above for $\mu$ in Erd\"os-R\'enyi networks
is a particular case of Eq.~(\ref{mu_inf}), since for these networks one has  
$\langle k^2 \rangle / \langle k \rangle = \langle k \rangle + 1$.
As indicated above, the number of SAWs on regular lattices scales for large
$n$ as $a_n \sim n^{\Gamma - 1} \mu^n$ \cite{pr91,so95}. 
For unclustered SF networks one has $a_n \sim \mu^n$, 
indicating that $\Gamma = 1$, the same exponent as
for regular lattices in $D > 4$.

For scale-free networks with $\gamma > 3$, both $\langle k \rangle$
and $\langle k^2 \rangle$ converge to finite values in the large-system
limit. Thus, for unclustered networks (triangle density $\nu = 0$) one can 
approximate the average values in Eq.~(\ref{mu_inf}) by those given in
Sec.~III.A for $\langle k \rangle_{\infty}$ and $\langle k^2 \rangle_{\infty}$, 
yielding
\begin{equation}
     \mu \approx k_0 \; \frac{\gamma - 2}{\gamma - 3} - 1   \;  .
\label{mu_inf2}
\end{equation}
For large $\gamma$ we recover the connective constant corresponding
to random regular networks ($k_i = k_0$ for all nodes), 
$\mu = k_0 - 1$, as for the Bethe lattice commented above. 

For $\gamma \leq 3$ the mean value $\langle k^2 \rangle$ diverges for 
$N \to \infty$, and the connective constant $\mu$ defined in Eq.~(\ref{mulim}) 
also diverges in the large-size limit \cite{he05a}.
In this case  we will consider a size-dependent connective constant
$\mu_N$ defined as
\begin{equation}
   \mu_N = \frac{\langle k^2 \rangle_N}{\langle k \rangle_N} - 1  \; .
\label{mu_n}
\end{equation}
Then, for unclustered SF networks with $\gamma = 3$ and $\nu = 0$, one has
\begin{equation}
   \mu_N  \approx  \frac12 k_0 \ln N   \; .
\label{mu_n3}
\end{equation}
For $2 < \gamma < 3$ and $\nu = 0$, $\mu_N$ behaves for large unclustered 
networks as
\begin{equation}
   \mu_N  \approx  k_0 \frac{\gamma - 2}{3 - \gamma}
       \left( \frac{\sqrt{N}}{k_0} \right)^{3 - \gamma}   \; .
\label{mu_n4}
\end{equation}

We emphasize that the expressions given in Eqs.~(\ref{mu_inf}) and 
(\ref{mu_n}) for the connective constants $\mu$ and $\mu_N$, 
as well as those presented in
this Section for SF networks with different values of $\gamma$, are
valid for unclustered networks (configuration model).
For clustered networks (in our case, triangle density $\nu > 0$),
those expressions are not valid because triangles introduce correlations
in the degrees of adjacent nodes.
In this case, one has to implement a different procedure to 
calculate the ratio $a_n / a_{n-1}$ necessary to obtain the connective
constant of clustered networks.

\section{Self-avoiding walks: analytical procedure}

In this section we present an analytical method to calculate the connective
constant in clustered scale-free networks, based on an iterative procedure
to obtain the average number of walks $a_n$ for increasing walk length $n$.
The ratio $a_n / a_{n-1}$ derived from this method converges fast for
the networks studied here (parameter $\gamma > 2$). 
In this procedure we take advantage of the fact that the number
of links $s_i$ and $t_i$ connected to a node $i$ are independent.

The number of $n$-step self-avoiding walks, $a_n$, can be written as 
\begin{equation}
    a_n = b_n + c_n  \; ,
\label{an}
\end{equation}
where $b_n$ and $c_n$ are the number of walks for which
step $n$ proceeds via an $s$-link or a $t$-link, respectively
(no matter the kind of links employed in the previous steps).
Note that the quantities $a_n$, $b_n$, and $c_n$, are average values for 
all possible starting nodes in the considered networks.
To simplify the notation,
we will call $\alpha = 2 \nu$, i.e., $\alpha$ is the average
number of $t$-links connected to a node, $\alpha = 2 \langle t \rangle$.
Moreover, we will call
\begin{equation}
   z = \langle s \rangle,  \hspace{1cm}  
         r = \frac{\langle s^2 \rangle}{\langle s \rangle}    \; .
\label{zr}
\end{equation}

To obtain $b_n$ and $c_n$ we will use the iteration formulas (for $n > 1$):
\begin{equation}
\left.
\begin{array}{lll}
b_n & = & (r - 1) \, b_{n-1} + z \, c_{n-1}     \\
c_n & = & \alpha \, b_{n-1} + (\alpha + 1) \, c_{n-1}  - \alpha \, a_{n-3} 
\end{array}
\right\}
\label{bncn}
\end{equation}
The initial conditions are: $b_1 = z$, $c_1 = \alpha$, 
 $a_0 =1$,  $a_{-1} = 0$.
In the first equation of (\ref{bncn}), $b_n$ is calculated from the number
of walks $b_{n-1}$ and $c_{n-1}$ of length $n-1$. 
If step $n-1$ goes on an $s$-link, then the prefactor is $r - 1$ 
to avoid returning on the previous link, as for unclustered SF networks with 
only $s$-links [see Eq.~(\ref{an1})].
If step $n-1$ proceeds on a $t$-link, then the prefactor of $c_{n-1}$ is $z$
(variables $s$ and $t$ are independent).

To calculate $c_n$ in the second equation of (\ref{bncn}) we have
contributions coming from $b_{n-1}$ and $c_{n-1}$, but there also appears a
third term with negative contribution corresponding to the self-avoiding 
condition, i.e., links associated to closing triangles are not allowed
(here we call ``closing a triangle'' to visit its three edges
in three successive steps of a walk). 
The first two terms on the r.h.s. are obtained from inputs of
step $n-1$. If step $n-1$ follows an $s$-link, then
the prefactor for $b_{n-1}$ is the mean number of $t$-links per node, 
i.e., $\alpha$. If step $n-1$ proceeds over a $t$-link, the prefactor 
for $c_{n-1}$ on the r.h.s. of the second equation is $\alpha + 1$.
This requires a comment.
Remember that in random networks the average degree of a randomly-chosen 
end node of a randomly selected edge is given by 
the ratio $\langle k^2 \rangle / \langle k \rangle$ \cite{do02,ne10},
as indicated above in connection to Eq.~(\ref{an1}).
Similarly, in our network of triangles, the average number of triangles,
$\langle t \rangle'$,
linked to a randomly-selected end node of a randomly taken $t$-link is
given by $\langle t^2 \rangle / \langle t \rangle$.
For the Poisson distribution of triangles we have
$\langle t \rangle = \nu$ and $\langle t^2 \rangle = \nu (\nu + 1)$,
so that $\langle t \rangle' = \nu + 1$. 
Hence, since one $t$-link was already visited in step $n-1$ (this refers
to the input associated to $c_{n-1}$), the number of 
$t$-links available for step $n$ is $2 (\nu + 1) - 1 = \alpha + 1$
(each triangle includes two available links).
 
The prefactor for $a_{n-3}$ in the second equation of (\ref{bncn}) is 
the average number of $t$-links corresponding to triangles that {\it close} 
at step $n$.
For an SAW of $n-3$ steps, there is an average number of $\nu$ triangles 
that may close after three more steps in step $n$.

Using both equations in (\ref{bncn}), one can derive the connective
constant $\mu$ from the asymptotic limit of the ratio $a_n / a_{n-1}$.
This is presented in Appendix A. 
We find that $\mu$ can be calculated by solving the system
[see Eqs.~(\ref{sigma1}), (\ref{mu2}), and (\ref{thetamu3})]
\begin{equation}
\left.
\begin{array}{l}
  \theta  \mu^3   -  (\theta \alpha + \theta + \alpha) \mu^2 +
          \alpha ( 1 + \theta ) =  0   \\
  \mu  -  z  \theta - r + 1  = 0
\end{array}
\right\}
\label{mutheta}
\end{equation}
where $\mu$ and $\theta$ are unknown variables.
$\mu$ is the connective constant defined above, and $\theta$ is 
the asymptotic limit of the ratio $c_n / b_n$ for large $n$.   
Both equations can be
combined to yield a quartic equation in $\mu$ with coefficients defined 
from the network parameters $z$, $r$, and $\alpha$.
For the networks considered here, this quartic equation has a single
positive solution, so that $\mu$ is univocally defined.
We note that the initial conditions $b_1$ and $c_1$ in the system 
(\ref{bncn}) are not relevant for the actual solution of the system.
This means that the connective constant derived from (\ref{bncn}) is
robust, in the sense that putting for $b_1$ and $c_1$ the values of
a particular node $i$ ($s_i$ and $t_i$) does not change the result
for $\mu$, which is a long-range characteristic of each network. 
Putting mean values for the starting conditions helps
to accelerate the convergence of the procedure.

\begin{figure}
\vspace{-0.9cm}
\includegraphics[width=8.0cm]{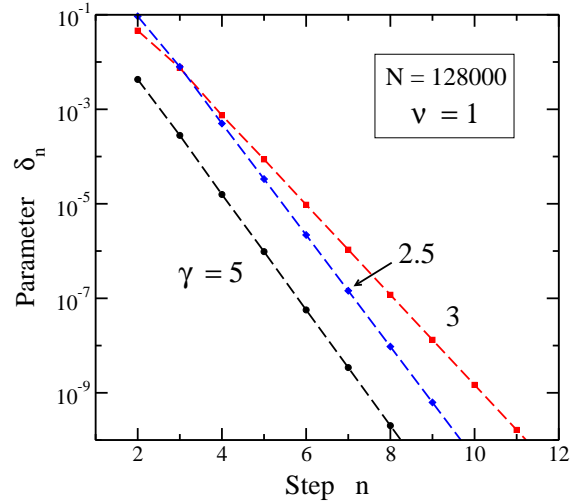}
\vspace{-0.5cm}
\caption{
Parameter $\delta_n$ as a function of the step number $n$ for clustered
scale-free networks with triangle density $\nu = 1$, size $N = 128,000$
nodes, and minimum degree $k_0 = 3$.
Symbols correspond to different values of the exponent $\gamma$:
5 (circles), 3 (squares), and 2.5 (diamonds).
}
\label{f4}
\end{figure}

In the case $\nu = 0$ (i.e., absence of triangles in unclustered networks,
$\alpha = 0$), one has, from Eqs.~(\ref{bncn}), 
$c_n^{\rm un} = 0$ for all $n$ and 
$b_n^{\rm un} = (r - 1) b_{n-1}^{\rm un}$ with 
$b_1^{\rm un} = \langle s \rangle$ (here the superscript `un' means
unclustered).  We thus recover the general expression for 
unclustered networks, Eq.~(\ref{an1}), and then
$\mu^{\rm un} = \langle s^2 \rangle / \langle s \rangle - 1$.

For $\gamma \leq 3$, we have defined a size-dependent connective constant
$\mu_N$. In this case, the limits to infinity presented above in this
section have no sense. However, the ratios $a_n / a_{n-1}$,
$b_n / b_{n-1}$, and $c_n / c_{n-1}$ considered here converge with our
present method for relatively low number of steps, $n \ll N$.
To analyze the convergence of the ratio $f_n = a_n / a_{n-1}$,
we define the parameter $\delta_n = |f_n - \mu_N | / \mu_N$.
In Fig.~4  we present in a logarithmic plot the parameter $\delta_n$ as 
a function of the step number $n$ for clustered scale-free networks 
with triangle density $\nu = 1$ and size $N = 128,000$. 
Symbols correspond to different values of the exponent $\gamma$:
5 (circles), 3 (squares), and 2.5 (triangles).
It appears that the couple of equations (\ref{bncn}) yields values of
$f_n$ that converge fast to the corresponding limit $\mu_N$ in a
similar way for networks with $\gamma < 3$ and $\gamma > 3$.
In the latter case, $\mu_N \to \mu_{\infty}$ for large $N$.

\section{Comparison with network simulations}

We have generated clustered scale-free networks by following the
procedure described in Sec.~II. For comparison, we have also constructed
unclustered networks, according to the configuration model, with 
the same degree distribution $P(k)$ than the clustered networks.
For both kinds of networks we have calculated the connective constant for
several values of the exponent $\gamma$, from the ratio $a_n / a_{n-1}$.
The results have been compared with those obtained by using the analytical
procedures described in Sec.~IV. We present these results in three
subsections, according to the behavior of the connective constant for
different values of $\gamma$.

\begin{figure}
\vspace{-0.9cm}
\includegraphics[width=8.0cm]{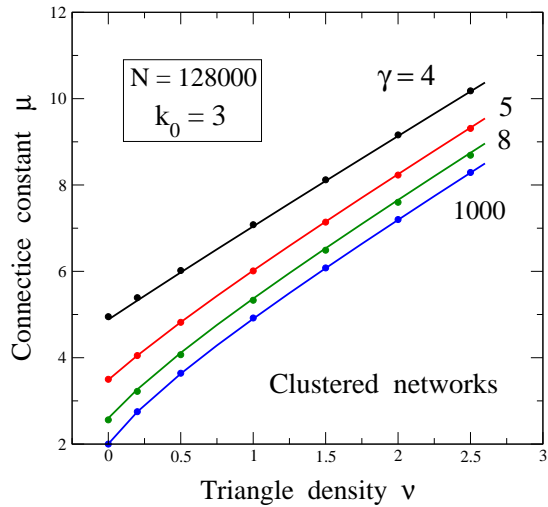}
\vspace{-0.5cm}
\caption{
Connective constant $\mu$ vs triangle density $\nu$
for clustered scale-free networks with several values of the exponent
$\gamma$:  From top to bottom $\gamma$ = 4, 5, 8, and 1000.
The networks size is $N$ = 128,000 and the minimum degree $k_0 = 3$.
Solid lines correspond to the analytical procedure described in
Sec.~IV, whereas symbols were derived from enumeration of SAWs
in clustered networks.  Error bars are less than the symbol size.
}
\label{f5}
\end{figure}

\subsection{Case $\gamma > 3$}

In this case the average value $\langle k^2 \rangle_N$ converges to
a finite value in the large-size limit, and therefore the connective
constant is well defined in this limit.
In Fig.~5 we display $\mu$ as a function of the
 triangle density $\nu$ for clustered scale-free networks with 
size $N = 128,000$, minimum degree $k_0 = 3$, and various values 
of the exponent $\gamma$. Symbols are data points derived from 
simulations for $\gamma$ = 4, 5, 8, and 1000. 
Here the large exponent $\gamma = 1000$ is equivalent to the limit
$\gamma \to \infty$, as in this case $s_i = k_0$ for all nodes.
Error bars in Fig.~5 are smaller than the symbol size.
Lines represent results obtained from the analytical method
presented above, and follow closely the data points obtained
from the network simulations. 
For $N = 128,000$ the finite-size effect in the connective constant
is almost inappreciable for networks with $\gamma > 3$.

For $\nu = 0$ (no triangles) the connective constant can be directly derived 
from the mean values $\langle s \rangle_N$ and $\langle s^2 \rangle_N$  given
in Eqs.~(\ref{smean_app}) and (\ref{smean2_app}), as
\begin{equation}
  \mu_N  =  \frac{\langle s^2 \rangle_N}{\langle s \rangle_N} - 1  \; ,
\end{equation}
as follows from Eq.~(\ref{an1}) for unclustered networks.
Then, for large $N$, $\mu_N$ converges to $\mu_{\infty}$ 
given in Eq.~(\ref{mu_inf2}).
For clustered networks with $\nu > 0$, the connective constant can be 
obtained by introducing the parameters $z$ and $r$ derived from
$\langle s \rangle_N$ and $\langle s^2 \rangle_N$ [see Eq.~(\ref{zr})] 
into the iterative equations (\ref{bncn}), or alternatively into
the system (\ref{mutheta}).
This gives for different values of $\gamma$ the lines shown in Fig.~5.

For unclustered networks with $\nu > 0$, $\mu_{\infty}$ is derived
from $\langle k \rangle_{\infty}$ and $\langle k^2 \rangle_{\infty}$
in Eqs.~(\ref{kmean}) and (\ref{k2mean}).
For a given network size ($N = 128,000$ in Fig.~5), $\mu_{\infty}$
converges to $2 \nu$ for large $\nu$, irrespective of the
exponent $\gamma$. This comes from the dominant terms in the ratio
$\langle k^2 \rangle_{\infty} / \langle k \rangle_{\infty}$,
i.e. $\langle k^2 \rangle_{\infty} \approx 4 \nu^2$ and
$\langle k \rangle_{\infty} \approx 2 \nu$ for $\nu \gg 1$.

The results for the connective constant $\mu_{\infty}$ of unclustered 
networks are very close to those corresponding to clustered networks,
and are not shown in Fig.~5. Both sets of results are in fact 
indistinguishable within error bars, which are smaller than the symbol size.
This is mainly due to the very low probability of
high-degree nodes in these networks.
This fact, which occurs for $\mu_{\infty}$ in networks
with $\gamma > 3$, does not happen for networks with $\gamma \leq 3$,
where results for $\mu_N$ appreciably differ for clustered and unclustered
networks (see below).

\begin{figure}
\vspace{-0.9cm}
\includegraphics[width=8.0cm]{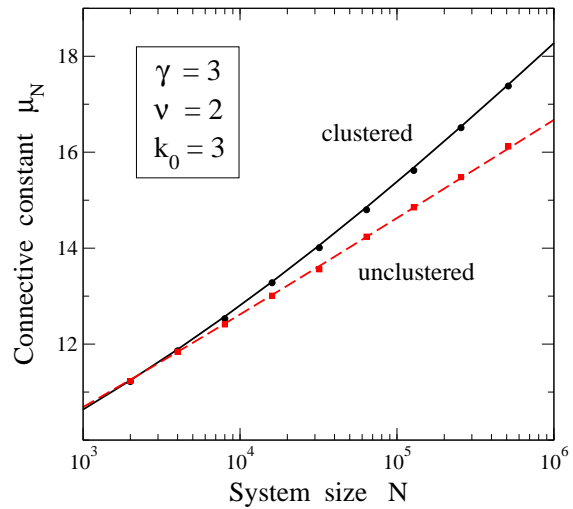}
\vspace{-0.5cm}
\caption{
Connective constant $\mu_N$ as a function of system size $N$
for scale-free networks with $\gamma = 3$ and triangle density
$\nu = 2$. The solid and dashed lines were obtained analytically
for clustered and unclustered (configuration model) networks, respectively.
Symbols correspond to results found from simulated scale-free networks.
Error bars are smaller than the symbol size.
}
\label{f6}
\end{figure}

\subsection{Case $\gamma = 3$}

For scale-free networks with $\gamma = 3$ the average value
$\langle k^2 \rangle_N$ diverges logarithmically with $N$, 
see Eq.~(\ref{smean2_app3}).
This behavior controls the asymptotic dependence of the connective
constant on system size.  In Fig.~6 we present the size-dependent $\mu_N$
vs $N$ for scale-free networks with $\gamma = 3$ and $\nu = 2$.
Shown are results for clustered and unclustered (configuration model)
networks. 
As in the case $\gamma > 3$ shown above, symbols are data points derived 
from SAWs in simulated networks, whereas lines were obtained from 
analytical calculations.
For small networks, $\mu_N$ is similar for both kinds of networks, and the 
connective constant corresponding to the clustered ones becomes appreciably 
larger as the system size increases.
For given $N$ and $\nu > 0$, we have $\mu_N^{\rm cl} > \mu_N^{\rm un}$.

For clustered networks (solid line), the procedure to calculate $\mu_N$ is 
the same as that employed above for $\gamma > 3$, using the iterative equations
(\ref{bncn}), and taking into account the adequate 
expressions for $\langle s \rangle_N$ and $\langle s^2 \rangle_N$ given in 
Eqs.~(\ref{smean_app}) and (\ref{smean2_app2}), respectively.
We cannot write an exact analytical expression for $\mu_N$ as a function of
$N$ and $\nu$ for clustered networks, but we can analyze its behavior 
for large $N$ and $\nu$ from the results of our calculations and 
network simulations.
For large $N$ and relatively small $\nu$, the r.h.s. of the first 
equation in the iterative system 
Eq.~(\ref{bncn}) giving $b_n$ is dominated by the contribution of $b_{n-1}$ 
because $r \gg z$. Then, $b_n / b_{n-1}$ converges to 
$\mu_N \approx \langle s^2 \rangle_N / \langle s \rangle_N \approx 
\frac12 k_0 \ln N$. 

\begin{figure}
\vspace{-0.9cm}
\includegraphics[width=8.0cm]{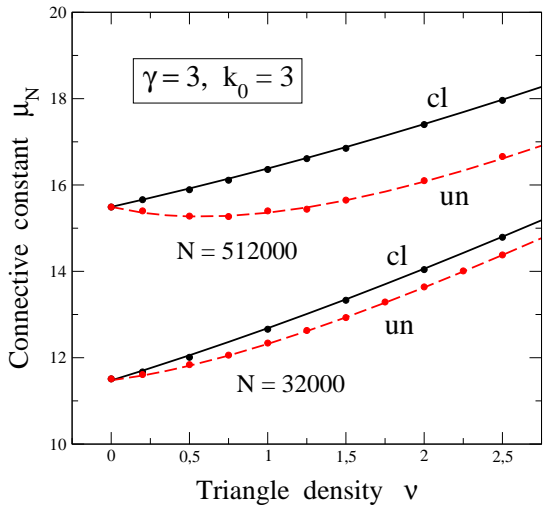}
\vspace{-0.5cm}
\caption{
Connective constant $\mu_N$ vs triangle density $\nu$
for scale-free networks with $\gamma = 3$ and $k_0 = 3$.
The data shown correspond to two different network size:
$N$ = 512,000 (top) and 32,000 (bottom).
Solid and dashed lines correspond to analytical calculations for
networks with and without triangles, respectively.
Symbols are results derived from network simulations.
Error bars are of the order of the symbol size.
`cl' and `un' refer to clustered and unclustered SF networks,
respectively.
}
\label{f7}
\end{figure}

For unclustered networks (dashed line in Fig.~6) $\mu_N$ is given by
Eq.~(\ref{mu_n}), valid for the configuration model.
Here $\langle k \rangle_N$ and $\langle k^2 \rangle_N$ are obtained from
average values corresponding to $s$-links and $t$-links separately, see
Eq.~(\ref{k2s2}).    Then, for large $N$ the ratio 
$\langle k^2 \rangle_N / \langle k \rangle_N$ is dominated by
the contribution of $s$-links, yielding
\begin{equation}
  \mu_N  \approx  \frac {\langle k^2 \rangle_N} {\langle k \rangle_N} 
      \approx  \frac {\langle s^2 \rangle_N} {\langle s \rangle_N + 2 \nu}
      \approx  \frac {k_0^2 \ln N} {2 (k_0 + \nu) }   \;  ,
\label{munk}
\end{equation}
which coincides with Eq.~(\ref{mu_n3}) for $\nu = 0$.
This means that for the data shown in Fig.~6 ($\nu = 2, k_0 = 3$), the slope
of the dashed line for large $N$ is given by 
$\partial \mu_N / \partial \ln N = 0.9$.
For SF networks with $\gamma = 3$ and $\nu = 0$, one has
$\mu_N \sim \frac12 k_0 \ln N$ for large $N$, Eq.~(\ref{mu_n3}), and
we find a slope $\partial \mu / \partial \ln N = 1.5$. 

In Fig.~7 we show $\mu_N$ for networks with $\gamma = 3$ as a function of
the triangle density $\nu$. We present results for clustered and unclustered
networks and two network sizes: $N$ = 32,000 and 512,000 nodes.
For both network sizes, $\mu_N^{\rm cl} > \mu_N^{\rm un}$ for $\nu > 0$.
For clustered networks with relatively large $N$, 
$\mu_N$ increases monotonically for rising $\nu$, with a slope
$d \mu / d \nu$ that grows converging to a value of 2 for large $\nu$.
This can be derived from Eq.~(\ref{theta1}), where the variable
$\theta$, defined in (\ref{thetan}), is the limit of the ratio $c_n / b_n$ 
for large $n$. Taking into account that $c_n$ and $b_n$ are the mean
number of SAWs ending in $t$-links and $s$-links, respectively,
one expects that $\theta$ increases as the triangle density 
(or the number of $t$-links) increases.  Thus, we have 
$\theta \lesssim \langle t \rangle / \langle s \rangle$, as can
be found from numerical solutions of Eq.~(\ref{mutheta}) or
Eq.~(\ref{theta1}).
Then, for large $\nu$, i.e. $\alpha \gg 1$, we have 
$\theta \gg 1$, $\mu_N \gg 1$, $\sigma = 1 / \mu_N \ll 1$,
and from Eq.~(\ref{mutheta}) we find that $\theta \to 2 \nu / z$.
Using again Eq.~(\ref{mutheta}), one has
$\mu_N = r + z \theta - 1$, with $r$ and $z$ independent of $\nu$,
so that $\partial \mu_N / \partial \nu = 
z \, \partial \theta / \partial \nu$, which converges to 2 for large $\nu$.

For unclustered networks (configuration model) we have for large $\nu$,
see Eq.~(\ref{mu_n}):
\begin{equation}
  \mu_N  =  \frac{\langle k^2 \rangle_N}{\langle k \rangle_N} - 1
           \approx \frac {k_0^2 \ln N} {2 \nu} + 2 \nu  \; ,
\label{munk3}
\end{equation}
and for a given $N$, $\partial \mu_N / \partial \nu \to 2$ 
for large $\nu$.
However, for large $N$ this convergence is slower than in the case of
clustered networks, as can be observed in Fig.~7.

Summarizing the results presented in this section for clustered networks 
with $\gamma = 3$, we find that for large $N$ or $\nu$, the leading 
contribution to the connective constant $\mu_N$ appears to be either 
a function $f(N)$ or another function $g(\nu)$, depending on the values of
both variables $N$ and $\nu$. This means that the behavior of $\mu_N$ can 
be described as $\mu_N \sim f(N) + g(\nu) + h(N,\nu)$, where we include a 
contribution $h(N,\nu)$ that becomes negligible for large $N$ or $\nu$.
For $\gamma = 3$ we have $f(N) \to \frac12 k_0 \ln N$ for large $N$ 
and $g(\nu) \to 2 \nu$ for large $\nu$.
Then, for $k_0 \ln N \gg \nu$, one has $\mu_N \approx \frac12 k_0 \ln N$,
whereas for $k_0 \ln N \ll \nu$, $\mu_N \approx 2 \nu$.
This indicates that one has a crossover from a parameter region where
the behavior of the connective constant is dominated by the scale-free
character of the degree distribution $P_{\rm sf}(s)$, to another region
where it is controlled by the cluster distribution (triangles).
This is not a sharp crossover, and in the intermediate region a simple
decomposition of $\mu_N$ into two independent contributions is 
not valid, as indicated above with the function $h(N,\nu)$.
For a given $N$, the crossover occurs for a triangle density 
$\nu_c \sim \frac14 k_0 \ln N$.
Thus, for $N = 5 \times 10^5$, $\nu_c \approx 10$, and for
$N = 10^4$, $\nu_c \approx 7$.

\begin{figure}
\vspace{-0.9cm}
\includegraphics[width=8.0cm]{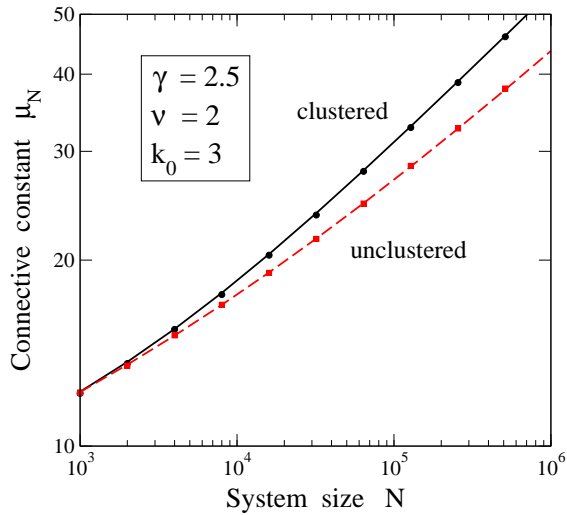}
\vspace{-0.5cm}
\caption{
Connective constant $\mu_N$ as a function of system size $N$
for scale-free networks with $\gamma = 2.5$ and triangle density
$\nu = 2$. The solid and dashed lines were obtained analytically
for clustered and unclustered networks, respectively.
Symbols correspond to results obtained from simulated
scale-free networks.
Error bars are less than the symbol size.
}
\label{f8}
\end{figure}

\subsection{Case $\gamma < 3$}

In this case the average value  $\langle k^2 \rangle_N$ scales as a power 
of the network size $N$, similar to $\langle s^2 \rangle_N$ in
Eq.~(\ref{smean2_app4}), and this dependence 
controls the behavior of the connective constant $\mu_N$.
In Fig.~8 we display $\mu_N$ as a function of $N$ for clustered and
unclustered SF networks with exponent $\gamma = 2.5$ and triangle density
$\nu = 2$. As in previous figures, symbols are data points obtained from
enumeration of SAWs in simulated networks, and
lines correspond to calculations following the procedure
described in Sec.~IV and Appendix A.
As for $\gamma = 3$ (see above), the connective constant for $\gamma = 2.5$
is similar for clustered and unclustered networks for relatively small $N$.
The results for both kinds of networks differ progressively as
$N$ increases, being larger the value for clustered networks, as in
Fig.~6 for $\gamma = 3$. Note, however, that the vertical scale in Fig.~8 
is logarithmic in contrast to Fig.~6, where it is linear.

For clustered networks with relatively large $N$ one has
$\mu_N  \approx  \langle s^2 \rangle_N / \langle s \rangle_N$, as
reasoned above for $\gamma = 3$. We find for $\gamma = 2.5$
a dependence $\mu_N \approx \sqrt{k_0} N^{1/4}$.
This agrees with the results shown in Fig.~8 (solid line), where
the derivative $\partial \ln \mu_N / \partial \ln N$ converges to
0.25 for $N \to \infty$. In fact this derivative rises for increasing $N$,
and for $N = 10^6$ it is already very close to its asymptotic limit.

\begin{figure}
\vspace{-0.9cm}
\includegraphics[width=8.0cm]{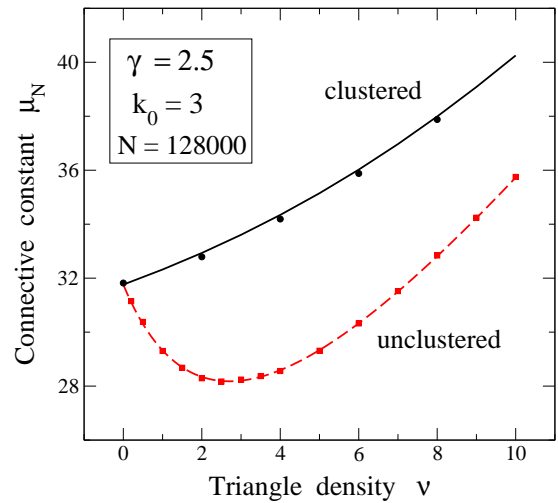}
\vspace{-0.5cm}
\caption{
Connective constant $\mu_N$ vs triangle density $\nu$ for clustered
SF networks with $\gamma = 2.5$: the solid line is the result of
the analytical model described in Sec.~IV, whereas solid circles are
data points derived from enumeration of SAWs
in simulated clustered networks.
The dashed line and solid squares correspond to unclustered networks
(configuration model). Error bars are less than the symbol size.
}
\label{f9}
\end{figure}

For unclustered networks with $\gamma = 2.5$ we have
\begin{equation}
  \mu_N  \approx  \frac {\langle k^2 \rangle_N} {\langle k \rangle_N}
      \approx  \frac {3 k_0^{3/2}} {3 k_0 + 2 \nu}  N^{1/4}   \; ,
\label{munk2}
\end{equation}
which coincides with Eq.~(\ref{mu_n4}) for $\nu = 0$.
Thus, for large $N$ the connective constant behaves in a similar way to
clustered networks, i.e. $\mu_N \approx A N^{1/4}$, but now  the prefactor
in Eq.~(\ref{munk2}) depends on the triangle density $\nu$.  
For the case presented in Fig.~8 ($\nu = 2$) we find $A = 1.20$ (large-$N$ 
limit of dashed line), in agreement with the results obtained from 
simulations of unclustered networks (solid squares). 
We note that the convergence to the power-law behavior $\mu_N \sim N^{1/4}$ 
is faster (occurs for smaller size $N$) for clustered 
than for unclustered networks.
In general, for $\gamma < 3$ one has
$\mu_N \sim  N^{(3-\gamma)/2}$,
as can be derived from the expressions for the mean values
$\langle s \rangle$ and $\langle s^2 \rangle$ given in Sec.~II.B.

In Fig.~9 we show the dependence of $\mu_N$ on triangle density $\nu$
for clustered and unclustered (configuration model) networks with
$\gamma = 2.5$ and system size $N = 128,000$ nodes. 
The connective constant for clustered networks increases monotonically with
the parameter $\nu$ and the slope $\partial \mu_N / \partial \nu$ 
grows and converges to 2 for large $\nu$. 
This happens in a way similar to the case $\gamma = 3$ discussed above
(Fig.~7).

For unclustered networks (solid squares and dashed line in Fig.~9) 
it is remarkable the decrease in $\mu_N$ for increasing $\nu$ close to
$\nu = 0$ for $N$ = 128,000, i.e., for small $\nu$ we have 
$\partial \mu_N / \partial \nu < 0$.   This is due to
the fact that, for growing $\nu$, the relative increase is larger for  
$\langle k \rangle_N$ than for $\langle k^2 \rangle_N$, so 
the ratio $\langle k^2 \rangle_N / \langle k \rangle_N$ decreases
for small $\nu$. 
This becomes more appreciable for larger network size. 
In the $\mu_N$ vs $\nu$ curve shown in Fig.~9 one has a minimum of $\mu_N$ 
for $\nu = 2.7$.
For higher triangle density $\nu$, the slope of the curve increases,
approaching the line corresponding to clustered networks.
For a given system size $N$, the quotient 
$\langle k^2 \rangle_N / \langle k \rangle_N$ yielding $\mu_N$ is 
dominated for large $\nu$ by the term 
$\langle t^2 \rangle = 4 \nu (\nu + 1)$ in the numerator
[see Eqs.~(\ref{k2s2}) and (\ref{k2mean})], 
and $\langle t \rangle = 2 \nu$ in the denominator, so 
$\partial \mu_N / \partial \nu \to 2$ in the large-$\nu$ limit.

Summarizing the results shown in this section for clustered networks
with $\gamma = 2.5$, we obtain that for large $N$ or $\nu$, 
the connective constant behaves as 
$\mu_N \sim \sqrt{k_0} N^{1/4} + g(\nu) + h(N,\nu)$, 
with $g(\nu) \sim 2 \nu$ for large $\nu$. 
As in the case $\gamma = 3$ discussed above, one has
in general a contribution $h(N,\nu)$ that becomes negligible in
the limits discussed here, and for which we do not have an analytical
expression.
Then, for $N \gg \nu^4$ we have 
$\mu_N \approx \sqrt{k_0} N^{1/4}$, and for $N \ll \nu^4$,
$\mu_N \approx 2 \nu$.
For a given $N$, there appears a crossover at a triangle density
$\nu_c \sim \frac12 \sqrt{k_0} N^{1/4}$. On one side, for $\nu < \nu_c$,
the behavior of the connective constant $\mu_N$ is dominated by
a power of $N$, corresponding to the scale-free degree distribution
of single links $P_{\rm sf}(s)$. On the other side, for 
$\nu > \nu_c$ , $\mu_N$ is controlled by the distribution of 
triangles $Q(t)$.
For $\gamma = 2.5$ and $N = 5 \times 10^5$ we have $\nu_c \approx 23$,
and for $N = 10^4$, $\nu_c \approx 9$.
In general, for $\gamma < 3$ the crossover takes place at a
triangle density $\nu_c$ given by
\begin{equation}
  \nu_c \sim  \frac12 \, \frac{\gamma-2}{3-\gamma} \, k_0^{\gamma-2}
	\,  N^{(3-\gamma)/2}  \; .
\label{nuc}
\end{equation}
Note that in this expression the exponent $(3-\gamma)/2$ is the same
as that of the connective constant for large $N$:
$\mu_N \sim  N^{(3-\gamma)/2}$.
In the parameter region where the behavior of $\mu_N$ is controlled
by the large-degree tail of the power law $P_{\rm sf}(s)$
(i.e., $\nu <\nu_c$), the connective constant  can be
described by an expression $\mu_N \sim N^\beta$, where $\beta$ is
a variable parameter that evolves from zero for small system size
to $(3-\gamma)/2$ for large $N$. Specifically, one can define 
$\beta$ as the logarithmic derivative $\beta = d \ln \mu_N / d \ln N$.
Thus, from the results shown in Fig.~8 for clustered networks
with $\gamma = 2.5$ we find a parameter $\beta$ in the interval
from 0.14 for $N = 10^3$ to 0.247 for $N = 7 \times 10^6$.
The latter value is close to the exponent $(3-\gamma)/2 = 0.25$
corresponding to a pure scale-free network.
The variable parameter $\beta$ is in principle regulated by the
function $h(N,\nu)$ mentioned above, but we do not know a precise 
expression for it.

\section{Conclusions}

 Self-avoiding walks are a suitable means to analyze the 
long-distance properties of complex networks.
We have studied the connective constant for clustered scale-free networks 
by using an iterative analytical procedure that converges in a few steps.
The results of these calculations agree well with those derived
from direct enumeration of SAWs in simulated networks.
These data have been compared with those corresponding to unclustered networks
with the same degree distribution $P(k)$.
For large unclustered networks, the number of SAWs rises with the number
of steps $n$ as 
$a_n/a_{n-1} \approx \langle k^2 \rangle / \langle k \rangle - 1$, 
but in the presence of clusters the ratio $a_n/a_{n-1}$ depends
explicitly on the distribution of both $s$ and $t$-links.

Our results can be classified into two different groups, depending on 
the exponent $\gamma$ of the power-law degree distribution
in this kind of networks.
Comparing clustered and unclustered networks, the conclusions obtained
for $\gamma > 3$ differ from those found for $\gamma \leq 3$.
For networks with $\gamma > 3$, one has a well-defined connective constant 
$\mu_{\infty}$ in the thermodynamic limit ($N \to \infty$).  
Adding motifs (here triangles) to the networks causes an increase
in $\mu$, mainly due to a rise in the average value $\langle k^2 \rangle$.
Nevertheless, we find a small numerical difference in $\mu$ for clustered and 
unclustered networks with the same degree distribution $P(k)$.

For $\gamma \leq 3$, the size-dependent connective constant $\mu_N$ 
is similar for clustered and unclustered networks with the same $P(k)$
when one considers small network sizes ($N \lesssim 10^3$).
This behavior changes for larger networks, where 
$\mu_N^{\rm cl} > \mu_N^{\rm un}$.
This difference is more apparent for decreasing $\gamma$, due to
the larger number of high-degree nodes.
For SF networks with $\gamma \leq 3$, $\mu_N$ increases with system size $N$, 
and diverges for $N \to \infty$.  
Depending on the values of the system size $N$ and triangle density $\nu$,
we find two regimes for the connective constant $\mu_N$. For a given
$N$ there appears a crossover from a region where $\mu_N$ is controlled
by the scale-free degree distribution $P_{\rm sf}(s)$ of single links
(for small $\nu$) to another parameter region where the behavior of
$\mu_N$ is dominated by the triangle distribution $Q(t)$ (large $\nu$).
The crossover between both regimes appears at a triangle density
$\nu_c$ such that $\nu_c \sim \frac14 k_0 \ln N$ for $\gamma = 3$
and $\nu_c \sim N^{(3-\gamma)/2}$ for $\gamma < 3$ [see Eq.~(\ref{nuc})].

The numerical results for clustered and unclustered scale-free networks
may change when different degree cutoffs are employed. This is
particularly relevant for $\gamma < 3$, since the dependence of $\mu_N$ 
on system size is important. 
In this respect, to avoid undesired correlations between degrees of 
adjacent nodes, we have employed here a degree cutoff 
$k_{\rm cut} = N^{1/2}$. 

Other probability distributions, different from the short-tailed Poisson 
type considered here for triangles, may be introduced to
change more strongly the long-degree tail in the overall degree
distribution $P(k)$. Thus, a power-law distribution for
the triangles can cause a competition between the exponents of both 
distributions (for $s$-links and $t$-links), which can change 
the asymptotic behavior of SAWs in such networks in comparison to
those presented here.

There are clear similarities between the asymptotic behavior of the
connective constant ($\mu_{\infty}$ or $\mu_N$) derived from SAWs and
the ferromagnetic-paramagnetic transition temperature for the Ising
model in this kind of networks \cite{he15}. 
This is related to the fact that both directly depend on the mean value
$\langle k^2 \rangle$, which changes with the exponent
$\gamma$ of the degree distribution. The extent of such similarities
in the case of clustered networks is an open question that should be 
further investigated.

\begin{acknowledgments}
This work was supported by Direcci\'on General de Investigaci\'on,
MINECO (Spain) through Grant FIS2015-64222-C2.
\end{acknowledgments}

\appendix

\section{Calculation of the connective constant}

In this section we present a derivation of the connective constant in
clustered networks
as the asymptotic limit of the ratio $a_n / a_{n-1}$. It is based
on the coupled iterative equations given in Eq.~(\ref{bncn}).
To find the values of $b_n$ and $c_n$ for large $n$, we define
\begin{equation}
 \theta_n = \frac{c_n}{b_n};  \hspace{2cm}
        \theta = \lim_{n \to \infty} \theta_n
\label{thetan}
\end{equation}
and
\begin{equation}
 \sigma_n = \frac{b_n}{b_{n+1}};  \hspace{2cm}
        \sigma = \lim_{n \to \infty} \sigma_n
\label{sigman}
\end{equation}
Then, from Eqs.~(\ref{bncn}), (\ref{thetan}), and (\ref{sigman}), we have
\begin{equation}
\theta_n = \frac { \alpha + \theta_{n-1} (\alpha + 1) -
             \sigma_{n-3} \sigma_{n-2} \alpha -
           \theta_{n-3} \sigma_{n-3} \sigma_{n-2} \alpha }
           { r - 1 + \theta_{n-1} z }  \; ,
\end{equation}
and taking the limit $n \to \infty$:
\begin{equation}
 \theta = \frac { \alpha + \theta (\alpha + 1) - \sigma^2 \alpha (1 + \theta) }
              {r - 1 + \theta z }  \; .
\label{theta1}
\end{equation}

Moreover, from Eq.~(\ref{bncn}) we have
\begin{equation}
   \frac{b_{n+1}}{b_n} = r - 1 +  \theta_n z   \; ,
\end{equation}
so that
\begin{equation}
  \frac{1}{\sigma} = \lim_{n \to \infty} \frac{b_{n+1}}{b_n} =
            r - 1 +  \theta  z   \; .
\label{sigma1}
\end{equation}
Eqs.~(\ref{theta1}) and (\ref{sigma1}) can be used to obtain the limits
$\theta$ and $\sigma$, as well as to find the simplified expression
\begin{equation}
  \frac{\theta}{\sigma} =
        \alpha + \theta (\alpha + 1) - \sigma^2 \alpha (1 + \theta)   \; .
\label{sigmatheta}
\end{equation}
Combining Eqs.~(\ref{sigma1}) and (\ref{sigmatheta}), one can eliminate
$\sigma$, which yields a quartic equation in $\theta$. 

The connective constant $\mu$ is defined as
\begin{equation}
   \mu = \lim_{n \to \infty} \frac{a_{n+1}}{a_n}   \; .
\label{mu}
\end{equation}
Taking into account that $a_n = b_n + c_n$ [Eq.~(\ref{an})], we have
\begin{equation}
 \mu = \lim_{n \to \infty}
  \frac{1 + \theta_{n+1}}{\sigma_n + \theta_n \sigma_n} = \frac{1}{\sigma}  \; .
\label{mu2}
\end{equation}
Using Eq.~(\ref{mu2}) we can rewrite Eq.~(\ref{sigmatheta}) as
\begin{equation}
  \theta  \mu^3   -  (\theta \alpha + \theta + \alpha) \mu^2 +
          \alpha ( 1 + \theta ) =  0  \; ,
\label{thetamu3}
\end{equation}
which can be used, along with Eq.~(\ref{sigma1}), to form the system 
of Eq.~(\ref{mutheta}) in the text.

We note that the asymptotic limit of both $b_{n+1}/b_n$ and
$c_{n+1}/c_n$ is also $\mu$.
In fact, from Eqs.~(\ref{sigman}) and (\ref{mu2}), we have
\begin{equation}
  \lim_{n \to \infty} \frac{b_{n+1}}{b_n} = \frac{1}{\sigma} = \mu   \; ,
\end{equation}
and
\begin{equation}
  \lim_{n \to \infty} \frac{c_{n+1}}{c_n} =
     \lim_{n \to \infty} \frac{c_{n+1}}{b_{n+1}} \frac{b_{n+1}}{b_n}
     \frac{b_n}{c_n} = \theta \mu \frac{1}{\theta} = \mu
\end{equation}

For $\gamma \leq 3$, we have defined in Sec~III a size-dependent connective 
constant $\mu_N$ . In this case, the limits to infinity presented in 
this Appendix have no sense. However, the ratios $a_n / a_{n-1}$,
$b_n / b_{n-1}$, and $c_n / c_{n-1}$ considered here converge with our
present method for relatively low number of steps, $n \ll N$, 
as indicated in Sec.~IV (see Fig.~4).


\end{document}